\documentclass[preprint]{aastex}
\usepackage{amsmath,amsfonts,amsthm,amssymb}
\usepackage{array}
\usepackage{subfigure}

\def\mf{\mathbf}

\def\mc{\mathcal}

\newcommand{\X}{\mathbf{\hat x}}
\newcommand{\Y}{\mathbf{\hat y}}
\newcommand{\Z}{\mathbf{\hat z}}
\newcommand{\intd}[1]{\ensuremath{\,\mathrm{d}#1}}

\begin{document}

\title{Parameter distributions of Keplerian orbits}		

\author{Dmitry Savransky}
\affil{Department of Mechanical and Aerospace Engineering\\ Princeton University, Princeton, NJ 08544}
\email{dsavrans@princeton.edu}

\author{Eric Cady}
\affil{Jet Propulsion Lab, California Institute of Technology, Pasadena, CA 91109}

\author{N. Jeremy Kasdin}
\affil{Department of Mechanical and Aerospace Engineering\\ Princeton University, Princeton, NJ 08544}

\begin{abstract}
Starting with just the assumption of uniformly distributed orbital orientations, we derive expressions for the distributions of the Keplerian orbital elements as functions of arbitrary distributions of eccentricity and semi-major axis.  We present methods for finding the probability density functions of the true anomaly, eccentric anomaly, orbital radius, and other parameters used in describing direct planetary observations.  We also demonstrate the independence of the distribution of phase angle, which is highly significant in the study of direct searches, and present examples validating the derived expressions.
\end{abstract}


\section{Introduction}
\citet{brown2004a} introduced the concept of completeness to study the selection effects introduced by observatory architectures on direct searches for sub-stellar companions.  Assuming distributions for semi-major axis and eccentricity of planetary orbits, Brown calculated the probability that a companion would fall outside the telescope's central obscuration during an observation of a star.  \citet{brown2005} subsequently expanded this concept to include the selection effects due to the photometric restrictions on observability introduced by telescope optics, and \citet{brown2009} demonstrated how completeness could be evaluated for indirect companion detection methods such as astrometry.  Completeness has also been extended to account for multiple observations of one star at different times \citep{brown2010new}, and has been utilized in mission analysis and development for a variety of proposed exoplanet observatories \citep{savransky2010,brown2009}.

The direct detection (imaging) completeness is evaluated by assuming that a companion will be observable if its angular separation from the star is greater than the observatory's inner working angle (IWA), and illuminated such that the difference in brightness between star and companion ($\Delta$mag) is below a threshold value, called the limiting $\Delta$mag, or $\Delta$mag$_0$.  The IWA represents the minimum angular separation between the telescope center-line and detectable objects on the sky.  It is determined by the size of a central obscuration, or the capability of adaptive optics systems to remove light from certain areas of the image plane, or the size and geometry of external occulting optics.  $\Delta$mag$_0$ represents the point where systematic errors produce unresolvable confusion between planet signal and background noise.  To calculate the completeness, probability distributions (or constant values) are assumed for planetary orbital elements and physical properties (see \S\ref{sec:orbit_defs}).  A large, equal number of samples is generated from each distribution, and the star-planet angular separation and $\Delta$mag are calculated for each set of samples.  When binned in a two-dimensional histogram, these generate a density function representing the probability that a planet drawn at random from the assumed population will have a given angular separation and $\Delta$mag.  Integrating over this density yields a cumulative distribution function (CDF), which can be used to determine the probability that an observatory with a given $\Delta$mag$_0$ and IWA, observing a specific star once,  will be able to detect a planet belonging to the assumed population.

The procedure described above is a Monte-Carlo sampling of a bivariate distribution function of non-independent arguments (since star-planet separation and $\Delta$mag are functions of the same parameters).  This means that to find any one point (or section) of the completeness distribution, it is necessary to sample it completely.  Because of the relatively high dimensionality of the initial parameter space and wide range of values certain parameters can take, complete sampling requires a large number of Monte-Carlo trials.  The first simulation in \citet{brown2005}, for example, includes 100 million samples, and it can be shown that certain low probability areas of the function are under-sampled.

Any alternate method of sampling completeness requires at least some knowledge of its density function.  In particular, Markov chain methods such as Metropolis-Hastings \citep{hastings1970monte} perform significantly better if the proposal distribution (a function used to `propose' new samples that are then either accepted or rejected) closely approximates the target distribution.  A special case of Metropolis-Hastings, known as Gibbs sampling, can be used to generate a sequence of samples from the joint distribution of two variables if their conditional probabilities are known.  Our goal is to derive the distribution functions of the arguments to the completeness functions.

We do so by starting with an ensemble of Keplerian orbits whose orientation is uniformly distributed in space, and deriving the distribution functions of these orbits' Keplerian parameters.  Rather than constraining the population of orbits, we make no assumptions as to the distribution of orbital semi-major axis and eccentricity.  This allows us to derive the completely general distribution functions presented in \S\ref{sec:pdfs}.  Building upon this, we further consider parameters related to direct planetary observations in \S\ref{sec:pdfsObs}, and make the discovery that the planetary phase angle (star-planet-observer angle) is independent of any of the orbital parameters.  This makes it possible to write distribution functions for quantities directly related to the two parameters of the direct detection completeness function.  It is important to note that, while the specific application explored here is direct imaging, the distributions derived in this paper are more broadly applicable to exoplanet studies in general.  For example, Keplerian fits are often employed in doppler spectroscopy surveys, making these derivations useful for inferring the true distributions of orbital parameters derived from radial velocity data sets.  Similarly, statistical analyses play an important role in other methods of exoplanet study, including transit photometry and microlensing surveys \citep{gould2010frequency}.

\section{Modeling Keplerian Orbits}\label{sec:orbit_defs}
In this section, we define the basic orbital parameters we will use throughout to describe companion orbits, and derive the relationships between these parameters and quantities that may be observed via direct planet searches.  We begin our derivation with three basic assumptions: First, that planetary orbits may be approximated as closed Keplerian orbits with negative specific orbital energy.  Second, that the distribution of orbital orientations with respect to an observer is uniform over a spherical volume. Finally, that any given star will have either one or no companions.  The first two assumptions allow us to describe any orbit with the parameter set $(a,e,\psi,\theta,\phi)$ where $a$ is the semi-major axis, $e$ is the eccentricity, and $\psi,\theta,\phi$ are Euler angles determining the orientation of the orbit in the observer's reference frame.  The position of the planet on its orbit at the time of observation is given by the true anomaly $\nu$.  The third assumption allows us to describe orbits in an exact fashion (via the Kepler solutions to the two-body problem), but means that the derived equations will fail to capture effects due to planet-planet interactions such as resonant orbits.

The observer's reference frame, $\mc I = (O,\X,\Y,\Z)$, is defined via a Cartesian coordinate set with the origin $O$ placed at the location of the observed star, and with the observer a distance $d$ from $O$ along the $-\Z$ axis.  A face-on orbit thus lies in the $(\X, \Y)$ plane, while an edge-on orbit lies in any plane orthogonal to the face-on one.  Starting with an orbit lying in the $(\Y, \Z)$ plane, with the planet's path counter-clockwise about the $\X$ axis, we apply a 3-1-3 rotation using the angle set $(\psi,\theta,\phi)$.  By the second assumption above, the angles $\psi$ and $\phi$ are uniformly distributed in $[0,2\pi]$, while $\theta$ is sinusoidally distributed in $[0,\pi)$--- i.e., $\theta$ is drawn from $\cos^{-1}(U([0,1]))$ where $U$ is a uniform distribution.

The position of the planet with respect to its parent star, at the time of observation, is thus:
\begin{equation} \label{eq:rpsdef}
\begin{array}{lcl}
\mf r_{p/s}  &=& 
\left[ \begin{matrix} \cos\phi &\sin\phi & 0 \\ -\sin\phi& \cos\phi & 0 \\ 0 & 0 & 1 \end{matrix}\right]
\left[ \begin{matrix} 1 & 0 & 0 \\ 0 & \cos\theta &\sin\theta \\0 & -\sin\theta & \cos\theta \end{matrix}\right]
\left[ \begin{matrix} \cos\psi &\sin\psi & 0 \\ -\sin\psi & \cos\psi & 0 \\ 0 & 0 & 1 \end{matrix}\right]
\left[ \begin{matrix} 0 \\ r\sin\nu \\r\cos\nu \end{matrix}\right]\\\\
&=&\left[
\begin{array}{c} r (\cos\nu  \sin\theta  \sin\phi +\sin\nu  (\cos\theta  \cos\psi  \sin\phi +\cos\phi
 \sin\psi )) \\
 r (\cos\nu  \cos\phi  \sin\theta +\sin\nu  (\cos\theta  \cos\phi  \cos\psi -\sin\phi
 \sin\psi )) \\
 r (\cos\theta  \cos\nu -\cos\psi  \sin\theta  \sin\nu )
\end{array}
\right]
\end{array}
\end{equation}
where $r$ is the distance between the planet and the star,
\begin{equation} \label{eq:rdef}
r = \Vert \mf r_{p/s}\Vert = \frac{a(1-e^2)}{e \cos(\nu) + 1} \,.
\end{equation}
Note that we have used left-handed rotations here to conform to conventions in the literature (see, for example, \citet{vinti}).  However, because of our second assumption and the inherent symmetries in the equations, all of the subsequent results can be reproduced if the starting formulation employs right-handed rotations instead.

The first two components of $\mf r_{p/s}$ can be found from the star-planet angular separation and a measurement of $d$, giving us the apparent separation (the distance of the planet from the star in the plane of the sky):
\begin{equation}  \label{eq:sdef}
s = \left\Vert \left[\begin{array}{ccc}1 & 0 & 0 \\ 0 & 1 & 0 \\ 0 & 0 & 0\end{array}\right]\mf r_{p/s}\right\Vert \,.
\end{equation}
The third component of $\mf r_{p/s}$ may also be observable if the detecting instrument can accurately measure the relative flux between the planet and its parent star, and some additional knowledge about the planet is assumed.  Following \citet{brown2005}, we write the ratio of fluxes between planet and star as:
\begin{equation} \label{eq:fluxRatio}
F_R \triangleq \frac{F_p}{F_s} = p\Phi(\beta) \left(\frac{R}{r}\right)^2 \quad \textrm{with} \quad \Delta\textrm{mag} = -2.5\log_{10}F_R \, ,
\end{equation}
where $p$ and $R$ are the planet's albedo and radius, respectively, $\Phi$ is the planet's phase function \citep{sobolev}, and $\beta$ is the star-planet-observer (phase) angle, given by:
\begin{equation}  
\beta = \cos^{-1}\left(\frac{\mf r_{p/o} \cdot  \mf r_{p/s} }{\Vert \mf r_{p/o} \Vert  \Vert\mf r_{p/s}\Vert}\right) \quad \textrm{where} \quad \mf r_{p/o} = \mf r_{p/s}
- \left[\begin{array}{ccc}0 & 0  & -d \end{array}\right] ^T \, .
\end{equation}
However, since $d \gg r $, we can make the approximation:
\begin{equation}  \label{eq:betadef}
\beta \approx \cos^{-1}\left(\frac{z}{r}\right)
\end{equation}
where $z$ is the $\Z$ axis component of $\mf r_{p/s}$.  For the closest star to Earth (approximately 1.3 pc away), this approximation produces errors of less than 0.001\%.  As the direct detection completeness is a function of only $s$ and $F_R$, equations (\ref{eq:sdef}) and (\ref{eq:fluxRatio}) are all that is needed to define the completeness.  Our eventual goal is  to find either the bivariate distribution function describing the direct detection completeness, or to find the marginal distribution functions of $s$ and $F_R$, which can be sampled to generate the completeness distribution.  To do so, we must first find expressions for the distributions of all of the parameters used in calculating these two quantities.

\section{Probability Density Functions of Orbital Parameters}\label{sec:pdfs}
As described in \S\ref{sec:orbit_defs}, the distributions of the Euler angles in the orbital parameter set are fixed by our assumption of uniform orbital orientation in space.  The remaining two parameters, $a$ and $e$, will be treated as unknowns, with probability density functions (PDFs) $f_{\bar{e}}(e)$ and $f_{\bar{a}}(a)$, respectively.  These distributions should ideally be generated via fits to observed data, which is an active area of current research. \citep{cumming2008,borucki2010kepler,hogg2010inferring}

To find the distribution of planetary positions upon their orbits, we begin by noting that we can relate $\nu$ to the eccentric anomaly ($E$) via the relationship:
\begin{equation} \label{eq:nutoE}
\tan\left(\frac{\nu}{2}\right) =\sqrt{\frac{1+e}{1-e}} \tan\left(\frac{E}{2}\right) 
\end{equation}
which, in turn, is related to the mean anomaly ($M$) by the Kepler equation:
\begin{equation} \label{eq:EtoM}
M = E - e\sin(E)
\end{equation}
The mean anomaly is proportional to the area covered by $\mf r_{p/s}$, measured from periapsis passage, and is thus linear in time.  A randomly timed observation therefore has equal probability of occurring at any value of $M$, so we let the mean anomaly be uniformly distributed in $[0, 2\pi]$.

Let $\bar{M}, \bar{e}$ and $\bar{\nu}$ be random variables with the distributions of mean anomaly, eccentricity and true anomaly, respectively, and define two functions: $\nu = g(M,e)$ and $M = h(\nu,e)$.  Following \citet{larson}, we can write the CDF for $\bar{\nu}$ as a marginalization of a conditional probability,
\begin{equation} 
F_{\bar{\nu}}(\nu) = P(g(\bar{M},\bar{e}) \le \nu) = \int_{-\infty}^{\infty} P(g(\bar{M},e) \le \nu | \bar{e} = e)f_{\bar{e}}(e)\, \mathrm{d}e=  \int_{0}^{1} P(\bar{M} \le h(\nu,e))f_{\bar{e}}(e)\, \mathrm{d}e\,,
\end{equation}
where the last step assumes independence between  $\bar{M}$ and $\bar{e}$, and that the orbits are closed so that $e \in [0,1)$.  We note that:
\begin{equation}\label{eq:Mprob}
P(\bar{M} \le h(\nu,e)) = \int_0^{h(\nu,e)} f_{\bar{M}}(M)\, \mathrm{d}M
\end{equation}
where $f_{\bar{M}}(M)$ is the PDF of $M$:
\begin{equation}\label{eq:Mdist}
f_{\bar M}(M) = \left\{
    \begin{array}{l l}
    \frac{1}{2 \pi} & M \; \in \; [0, 2\pi) \\
    0 & \mathrm{else}
    \end{array} \right. \,.
\end{equation}
Therefore the PDF of $\nu$ is:
\begin{equation}\label{eq:nupdf}
f_{\bar{\nu}}(\nu) =  \frac{\mathrm{d}}{\mathrm{d}\nu}F_{\bar{\nu}}(\nu) =  \frac{1}{2\pi} \int_{0}^{1} \frac{\partial h}{\partial \nu}f_{\bar{e}}(e)\, \mathrm{d}e  =  \frac{1}{2\pi} \int_{0}^{1} \frac{\left(1-e^2\right)^\frac{3}{2}}{\left(1+e\cos\nu\right)^2} f_{\bar{e}}(e)\, \mathrm{d}e \, ,
\end{equation}
where $h$ is given by equations (\ref{eq:nutoE}) and (\ref{eq:EtoM}).

Following the same procedure, but without substituting equation (\ref{eq:nutoE}) into equation (\ref{eq:EtoM}), we find that the PDF of the eccentric anomaly is:
\begin{equation} \label{eq:Edist}
f_{\bar{E}}(E) =  \int_{0}^{1} \frac{\partial }{\partial E}\left( E - e\sin(E) \right)f_{\bar{M}}(M)f_{\bar{e}}(e)de = \frac{1}{2\pi}\int_{0}^{1} \left(1-e\cos E\right) f_{\bar{e}}(e)\, \mathrm{d}e \,.
\end{equation}
Similarly, if we rewrite the Kepler equation as:
\begin{equation}
M = \cos^{-1}\left(\frac{a - r}{ea}\right) - e\sqrt{1 - \frac{(a-r)^2}{(ea)^2}}
\end{equation}
and assume independence between $\bar a$ and $\bar e$, we can write the PDF for the orbital radius as:
\begin{equation}\label{eq:rpdf}
f_{\bar{r}}(r) = \frac{1}{\pi}\int_{0}^{\infty} \int_{0}^{1} \frac{r}{a\sqrt{(ae)^2 - (a-r)^2}}f_{\bar{e}}(e) \, \mathrm{d}e \, f_{\bar{a}}(a)\, \mathrm{d}a \,.
\end{equation}
Equations (\ref{eq:nupdf}), (\ref{eq:Edist}), and (\ref{eq:rpdf}) are the most complete descriptions possible for the distributions of the orbital position and anomaly, without assuming specific distributions for the eccentricity and semi-major axis.  In \S\ref{sec:validate} we present examples where functions are selected for $f_{\bar e}(e)$ and $f_{\bar a}(a)$ and algebraic forms are calculated for $f_{\bar \nu}(\nu)$ and $f_{\bar r}(r)$.

\section{Probability Density Functions of Observed Quantities}\label{sec:pdfsObs}
We derived the distribution functions of the true anomaly and orbital radius by assuming independence between the Keplerian orbital elements from which the anomaly and radius are calculated.  However, the quantities observed by direct searches (i.e., $s$ and $F_R$) are nonlinear functions of the Keplerian elements and $r$, so we cannot make the assumption of independence between their functional arguments for either of these.  Fortunately, inspection of the phase angle reveals something interesting.  Using the approximation in equation (\ref{eq:betadef}), we can write:
\begin{equation}
\beta \approx \cos^{-1}\left(\cos\theta  \cos\nu -\cos\psi  \sin\theta  \sin\nu \right) 
\end{equation}
from which we define a new variable,
\begin{equation} \label{eq:cosBeta}
x \triangleq \cos\beta = \cos\theta  \cos\nu -\cos\psi  \sin\theta  \sin\nu \,.
\end{equation}
If it can be shown that $x$ is uniformly distributed for all distributions of eccentricity, then this would prove that $\beta$ is sinusoidally distributed regardless of the distribution of any other orbital parameter, given only our assumption of uniform orbital orientation.  We do so by considering the characteristic function of $x$,
\begin{equation}
\varphi_{\bar{x}}(t) =  \mathbb{E}\left(e^{i t \bar{x}}\right) = \int_{-\infty}^{\infty} \int_{-\infty}^{\infty} \int_{-\infty}^{\infty} e^{i t x} f_{\bar\nu}(\nu) f_{\bar\psi}(\psi) f_{\bar\theta}(\theta)  \intd{\psi} \intd{\theta} \intd{\nu} \,,
\end{equation}
where $\mathbb{E}$ is the expectation value, and showing that it is equivalent to the characteristic function of a uniformly distributed variable.

The assumption of uniform orbital orientation yields:
\begin{eqnarray}
f_{\bar\psi}(\psi) &=& \left\{
    \begin{array}{l l}
    \frac{1}{2 \pi} & \psi \; \in \; [0, 2\pi) \\
    0 & \mathrm{else}
    \end{array}
    \right. \label{eq:psipdf}\\
f_{\bar\theta}(\theta) &=& \left\{
    \begin{array}{l l}
    \frac{\sin{\theta}}{2} & \theta \; \in \; [0, \pi) \\
    0 & \mathrm{else}
    \end{array}
        \right. \label{eq:thetapdf}
\end{eqnarray}
so that the characteristic function of $x$ becomes:
\begin{equation}
\varphi_{\bar{x}}(t)  = \frac{1}{4 \pi}\int_{-\infty}^{\infty} \int_{0}^{\pi} \int_{0}^{2 \pi}  e^{i t (\cos{\theta}\cos{\nu}- \cos{\psi}\sin{\theta}\sin{\nu})}\intd{\psi} \sin{\theta} \intd{\theta}\, f_{\bar\nu}(\nu) \intd{\nu}
\end{equation}
Performing the integral over $\psi$,
\begin{equation}
    \varphi_{\bar{x}}(t)  = \frac{1}{2}\int_{-\infty}^{\infty} \int_{0}^{\pi}  e^{i t \cos{\theta}\cos{\nu}}  J_{0}\left(t \sin{\theta}\sin{\nu}\right)  \sin{\theta} \intd{\theta}\, f_{\bar\nu}(\nu) \intd{\nu}
\end{equation}
where $J_0$ is the zeroth-order Bessel function of the first kind.
We next perform the integral over $\theta$ using Gegenbauer's finite integral (see equation 12.14(1) in \citet{watson1944treatise}):
\begin{equation}
    \int_0^{\pi}  e^{i t \cos{\theta}\cos{\nu}} J_{b-\frac{1}{2}}\left(t \sin{\theta}\sin{\nu}\right) C_a^b{(\cos{\theta})} \sin^{b+\frac{1}{2}}{\theta} \intd{\theta} = \sqrt{\frac{2 \pi}{t}} i^a \sin^{b-\frac{1}{2}}{\nu} \, C_a^b{(\cos{\nu})} J_{a+b}(t)
\end{equation}
where $C_a^b$ is a Gegenbauer polynomial.  Choosing $a = 0$ and $b = \frac{1}{2}$ yields $C_a^b(x) = 1$ and the integral becomes:
\begin{equation}
    \int_0^{\pi} e^{i t \cos{\theta}\cos{\nu}} J_{0}\left(t \sin{\theta}\sin{\nu}\right) \sin{\theta} \intd{\theta} = \sqrt{\frac{2 \pi}{t}} J_{\frac{1}{2}}(t)
\end{equation}
Since the half-order Bessel function is defined as:
\begin{equation}
    J_{\frac{1}{2}}(t) = \sqrt{\frac{2}{\pi t}}\sin t
\end{equation}
the characteristic function becomes:
\begin{align}
 \varphi_{\bar{x}}(t)  &= \frac{1}{2} \int_{-\infty}^{\infty} \sqrt{\frac{2 \pi}{t}} \sqrt{\frac{2}{\pi t}}\sin t f_{\bar\nu}(\nu) \intd{\nu} \\
&= \frac{\sin t}{t} \int_{-\infty}^{\infty} f_{\bar\nu}(\nu) \intd{\nu}
\end{align}
By definition,
\begin{equation}
\int_{-\infty}^{\infty} f_{\bar\nu}(\nu) \intd{\nu} = 1 
\end{equation}
and thus,
\begin{equation} \label{eq:xChar}
 \varphi_{\bar{x}}(t) = \frac{\sin t}{t}
\end{equation}

If we define a random variable $\bar y \sim U([-1,1])$, its characteristic function will be:
\begin{align}
 \varphi_{\bar{y}}(t) =  \mathbb{E}\left(e^{i t \bar y}\right) &= \int_{-\infty}^{\infty} e^{i t y} f_{\bar y}(y) \intd{y} \nonumber \\
&= \frac{1}{2}\int_{-1}^{1} e^{i t y} \intd{y} = \frac{1}{2}\left(\frac{e^{i t }}{i t} - \frac{e^{-i t }}{i t}\right) \label{eq:uniformChar} \\ 
&= \frac{\sin t}{t} \nonumber
\end{align}
As equations (\ref{eq:xChar}) and (\ref{eq:uniformChar}) are identical, $\bar{x}$ and $\bar{y}$ have the same characteristic function, and thus must have the same underlying distribution.  Since $\bar x$ is uniformly distributed on $[-1, 1]$, $\beta$ is sinusoidally distributed in $[0,\pi)$ for any population of uniformly oriented orbits, independent of the specific distribution of $\bar e$ and $\bar a$.  This finding greatly simplifies our derivation of the distribution functions of the flux ratio and apparent separation.

Returning now to the definition of flux ratio in equation (\ref{eq:fluxRatio}), and assuming that we are interested in a particular planet type such that $pR^2$ is approximately constant, we see that the flux ratio is a function of the product of two variables, $m = \Phi(\beta)$ and $n = r^{-2}$.  If we define a random variable $\bar k \triangleq \bar m \bar n$, then it can be shown that:
\begin{equation}\label{eq:distmult}
f_{\bar{k}}(k) = \int_{-\infty}^\infty \frac{1}{n} f_{\bar{m}}\left(\frac{k}{n}\right)f_{\bar{n}}(n)\, \mathrm{d}n
\end{equation}
if $\bar m$ and $\bar n$ are independent. \citep{larson}  As we have demonstrated the independence of $\bar \beta$ from $\bar r$, equation (\ref{eq:distmult}) is applicable.
Using equation (\ref{eq:rpdf}), the CDF of $\bar n$ is given by:
\begin{align}
F_{\bar n}(n) &= P(\bar{r}^{-2} \le n) = P\left(-\frac{1}{\sqrt{n}} \le \bar{r} \le \frac{1}{\sqrt{n}}\right) \nonumber\\
&= \int_{-1/\sqrt{n}}^{1/\sqrt{n}} \frac{1}{2\pi}\int_{0}^{\infty} \int_{0}^{1} \frac{r}{a\sqrt{(ae)^2 - (a-r)^2}}f_{\bar{e}}(e) \, \mathrm{d}e \, f_{\bar{a}}(a)\, \mathrm{d}a \, \mathrm{d}r \\
&= \left. \frac{1}{2\pi}\int_{0}^{\infty} \int_{0}^{1} \sqrt{(ae)^2 - (a-r)^2} + 2 a \tan^{-1}\left(\sqrt{\frac{ae + a - r}{ae - a + r}}\right)\right|_{-1/\sqrt{n}}^{1/\sqrt{n}} f_{\bar{e}}(e) \, \mathrm{d}e \, \frac{f_{\bar{a}}(a)}{a}\, \mathrm{d}a \nonumber \,.
\end{align}
The PDF of $n$ is thus:
\begin{equation}
f_{\bar n}(n) = \frac{1}{4\pi}\int_{0}^{\infty} \int_{0}^{1}
n^{-\frac{3}{2}}\left[ \left(C_0 +2a\sqrt{n} - 1\right)^{-\frac{1}{2}} -\left(C_0 - 2a\sqrt{n} - 1\right))^{-\frac{1}{2}}\right] 
f_{\bar{e}}(e) \, \mathrm{d}e \, \frac{f_{\bar{a}}(a)}{a}\, \mathrm{d}a
\end{equation}
where $C_0 = a^2(e^2 - 1)n$.  The PDF of $\bar m$ is dependent on the exact form of $\Phi$, but can also be evaluated analytically when $\Phi$ is invertible.  In those cases,
\begin{equation}\label{eq:mpdf}
f_{\bar m}(m) = f_{\bar \beta}(\Phi^{-1}(m)) \left| \frac{ \mathrm{d}}{ \mathrm{d} m} \Phi^{-1}(m) \right| = 
\left\{
    \begin{array}{c l}
    \frac{1}{2} \sin\left(\Phi^{-1}(m)\right) \left| \frac{ \mathrm{d}}{ \mathrm{d} m} \Phi^{-1}(m) \right|  \quad &  \Phi^{-1}(m) \; \in \; [0, \pi) \\
    0 & \mathrm{else}
    \end{array} \right.
\end{equation}

This formulation is still problematic when dealing with transcendental functions such as the commonly used phase function of a Lambertian sphere \citep{sobolev}, 
\begin{equation}\label{eq:lambertPhase}
\pi \Phi(\beta) = \sin\beta + (\pi - \beta)\cos\beta \, ,
\end{equation}
but, since the domain of the function is restricted to $\beta \in [0,\pi]$, the range of $\Phi$ is single-valued ($\in [1, 0]$), and the function is invertible.  Expanding about $\pi/2$, equation (\ref{eq:lambertPhase}) becomes:
\begin{equation}\label{eq:lambertPhaseSeries}
\Phi(\beta) = \frac{1}{\pi}  + \sum_{k=1}^{\infty} \alpha_k \left(\beta - \frac{\pi}{2}\right)^k \quad\textrm{where}\quad \alpha_k = \frac{d_k}{2 k!}\left(\frac{2 (k-1)}{\pi}\right)^{\frac{d_k + d_{k+1}}{2 d_k}}
\end{equation}
and $d_{1:3} = -1,1,1$, with $d_k = d_{k-1}d_{k-2}d_{k-3}$ for $k > 3$, such that you get the series $d_i = \{-1,1,1,-1,-1,1,1,-1,-1\ldots\}$.  Following \citet{morese1953methods}, we can express the inverse function as the series:
\begin{equation}
\beta = \Phi^{-1}(w) = \frac{\pi}{2}  + \sum_{k=1}^{\infty} b_k \left(w - \frac{1}{\pi}\right)^k 
\end{equation}
where
\begin{equation}
b_k = \frac{1}{k \alpha_1^k} \sum_{x \in X} (-1)^{\vert x \vert} \prod_{r=1}^{\vert x\vert}(k-1+r)  \prod_{i=1}^{k-1}\frac{\left(\alpha_{i+1}/\alpha_1\right)^{x_i}}{x_i!}
\end{equation}
and the space $X$ of sets $x$ is defined as:
\begin{equation}\label{eq:Xspacedef}
X \triangleq \left\{x \in \mathbb{N}^{k-1} : \sum_{i=1}^{k-1} i x_i = k-1\right\} \quad\textrm{with}\quad \vert x\vert \triangleq \sum_i x_i \,.
\end{equation}
The inverse series converges to machine double-precision in a finite number of terms, except at the endpoints (0 and 1), which themselves do not need to be computed as they map exactly to $\beta = \pi$ and $\beta = 0$, respectively.

Since $\bar F_R = pR^2 \bar k$, equations (\ref{eq:distmult}) - (\ref{eq:Xspacedef}) allow us to write:
\begin{equation}\label{eq:FRpdf}
f_{\bar F_R}(F_R) = \frac{1}{pR^2}\int_{-\infty}^\infty \frac{f_{\bar{n}}(n)}{n} \cos\left(\sum_{k=1}^{\infty} b_k \left(\frac{F_R}{npR^2} - \frac{1}{\pi}\right)^k\right)\left|\sum_{k=1}^\infty k b_k \left(\frac{F_R}{npR^2} - \frac{1}{\pi}\right)^{k-1} \right| \, \mathrm{d}n \,,
\end{equation}
which represents the PDF of the planet-star flux ratio for a population of planets with constant $pR^2$ values (i.e., Earth-twins) modeled as Lambertian reflectors.

Finally, we consider the apparent separation.  The same approximation used in equation (\ref{eq:betadef}) allows us to write:
\begin{equation}
s = r \sin \beta \,.
\end{equation}
If we let $\bar l = \sin\bar\beta$, following equation (\ref{eq:mpdf}):
\begin{equation}\label{eq:lpdf}
f_{\bar l}(l) = f_{\bar \beta}(\sin^{-1}(l)) \left| \frac{ \mathrm{d}}{ \mathrm{d} l} \sin^{-1}(l) \right| = 
\left\{
    \begin{array}{l l}
    \frac{l}{\sqrt{1 - l^2}} & l \; \in \; [0,1) \\
    0 & \mathrm{else} \end{array}\right.
    \end{equation}
Returning to equations  (\ref{eq:rpdf}) and (\ref{eq:distmult}) and now letting $\bar m = \bar r$ and $\bar n = \bar l$, we have:
\begin{equation}\label{eq:spdf}
f_{\bar s}(s) = \frac{1}{\pi}  \int_{0}^1 \int_{0}^{\infty} \int_{0}^{1} \frac{s}{a\sqrt{\left(1 - l^2\right)\left[(ael)^2 - (al-s)^2\right]}}f_{\bar{e}}(e) \, \mathrm{d}e \, f_{\bar{a}}(a)\, \mathrm{d}a \, \mathrm{d}l  \, ,
\end{equation}
which represents the probability density function of the apparent separation.  We now have expressions for the distribution functions of both of the observed quantities of direct imaging planet searches.  While the expressions are quite complex, they are numerically integrable, and can be used to check and improve the efficiency of the sampling of the completeness function.   With an assumed $f_{\bar e}(e)$ and $f_{\bar a}(a)$, equations (\ref{eq:FRpdf}) and (\ref{eq:spdf}), via marginalization and joint sampling, can directly produce the completeness distribution.

\section{Validation of Derived Distributions}\label{sec:validate}
As a check on the derived distribution functions for the true anomaly and orbital radius, we first consider the simplest possible case, where both eccentricity and semi-major axis are uniformly distributed:
\begin{eqnarray}
f_{\bar e}(e) &=& \left\{
    \begin{array}{l l}
   1 & e \; \in \; [0, 1] \\
    0 & \mathrm{else}
    \end{array}
    \right. \label{eq:euniformpdf}\\
f_{\bar a}(a) &=& \left\{
    \begin{array}{l l}
    (a_{max} - a_{min})^{-1} &a \; \in \; [a_{max}, a_{min}] \\
    0 & \mathrm{else}
    \end{array}
        \right. \label{eq:auniformpdf}
\end{eqnarray}
Equation (\ref{eq:nupdf}) then becomes:
\begin{align} \label{eq:nupdfuniform}
f_{\bar{\nu}}(\nu) &=  \frac{1}{2\pi} \int_{0}^{1} \frac{\left(1-e^2\right)^\frac{3}{2}}{\left(1+e\cos\nu\right)^2}\, \mathrm{d}e  \nonumber \\
&= -{}_2F_1(1,2;7/2,\cos^2\nu)\frac{\cos\nu}{5\pi} - \frac{3}{16 \cos^4\nu}\left(4\sin\nu + \cos 2\nu - 3\right)
\end{align}
where ${}_2F_1$ is the Gaussian hypergeometric function.
Equation (\ref{eq:rpdf}) becomes:
\begin{align}
f_{\bar{r}}(r) =\frac{1}{\pi a \left(a_{max} - a_{min}\right)} & \left[
2 r \log\left(C_1 + \sqrt{r}\right) + a\log\left(a - r\right) - 2 r \log\left(C_2 - \sqrt{r - a}\right) \right. \nonumber\\
& {} + ia\log\left(\frac{4\sqrt{r}\left(2 C_2 + C_1 + i\sqrt{r}\right)}{C_2 + i\sqrt{r}}\right)- 2a\log\left(-2\left(r + \sqrt{r}C_1\right)\right) \label{eq:rpdfuniform} \\
& \left.\left.{}+ ia\log\left(\frac{4\sqrt{r}\left(-2 C_2 + C_1 + i\sqrt{r}\right)}{C_2 - i\sqrt{r}}\right)
\right]\right|_{a_{min}}^{a_{max}} \nonumber
\end{align}
where $C_1 = \sqrt{2a - r}$ and  $C_2 = \sqrt{a - r}$.

\begin{figure}[ht]
\centering
\subfigure[$f_{\bar{\nu}}$]{\includegraphics[width=0.49\columnwidth]{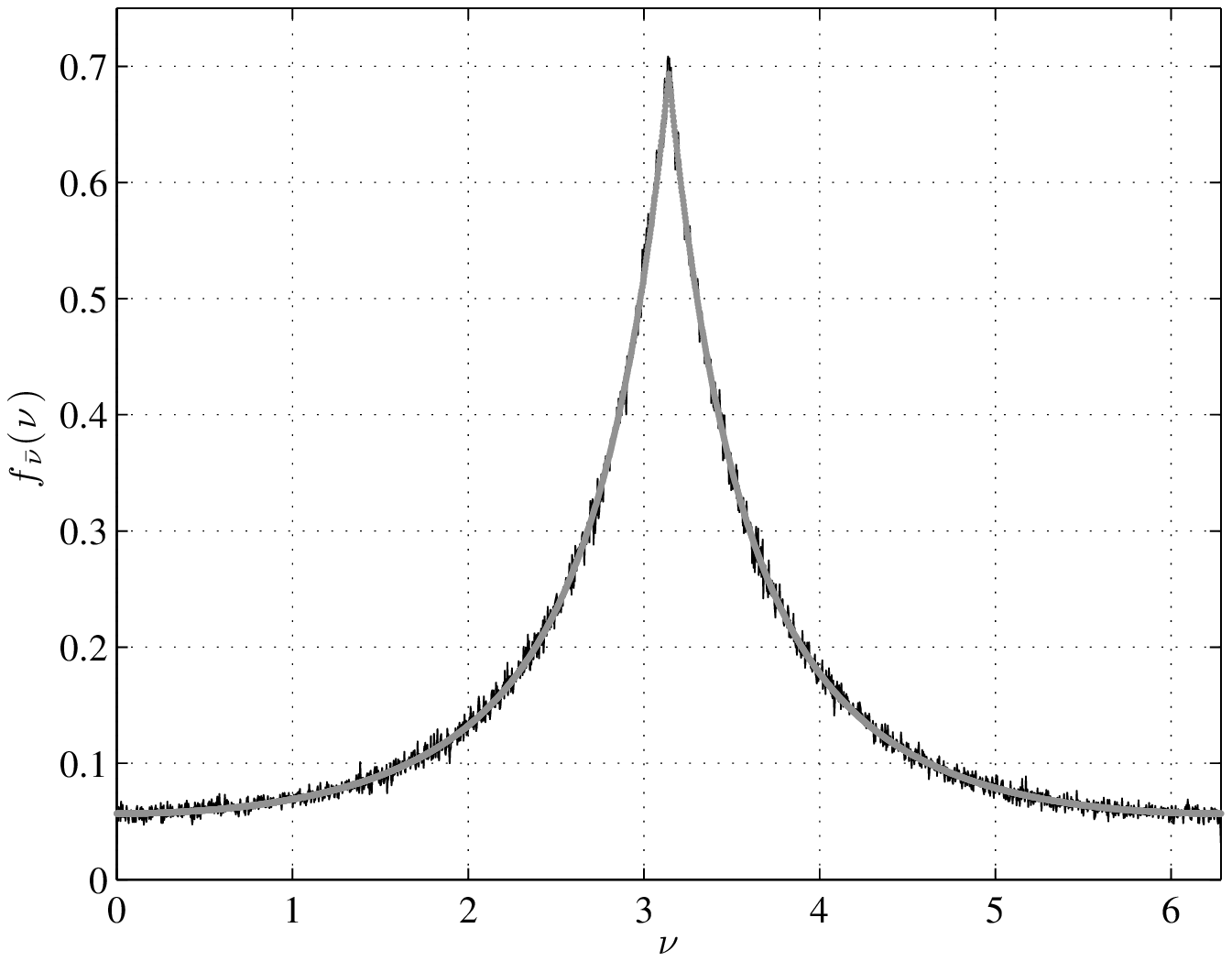}\label{fig:compDistsa}}
\subfigure[$f_{\bar{r}}$, uniform $a$]{\includegraphics[width=0.49\columnwidth]{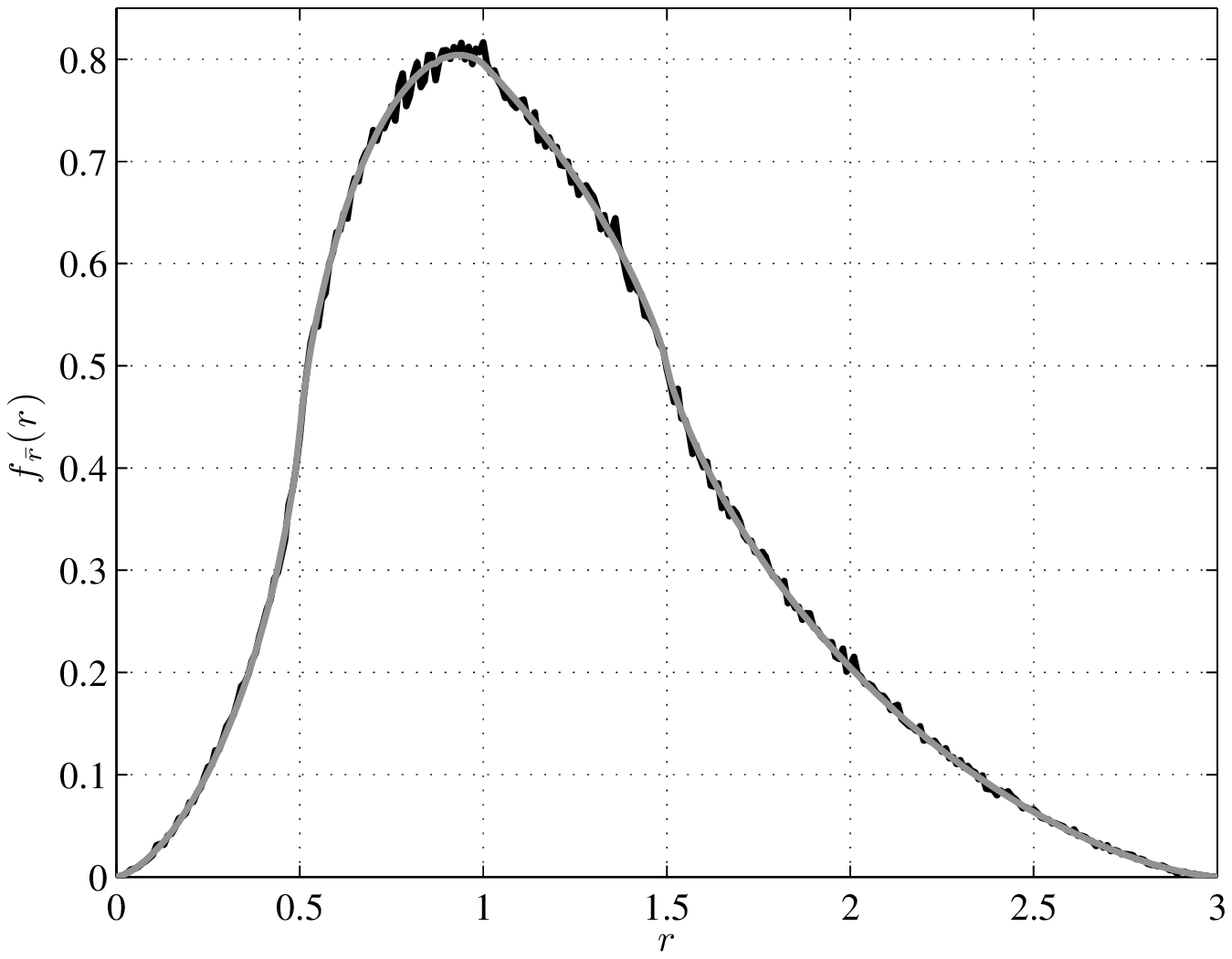}\label{fig:compDistsb}}
\subfigure[$f_{\bar{r}}$, log $a$]{\includegraphics[width=0.49\columnwidth]{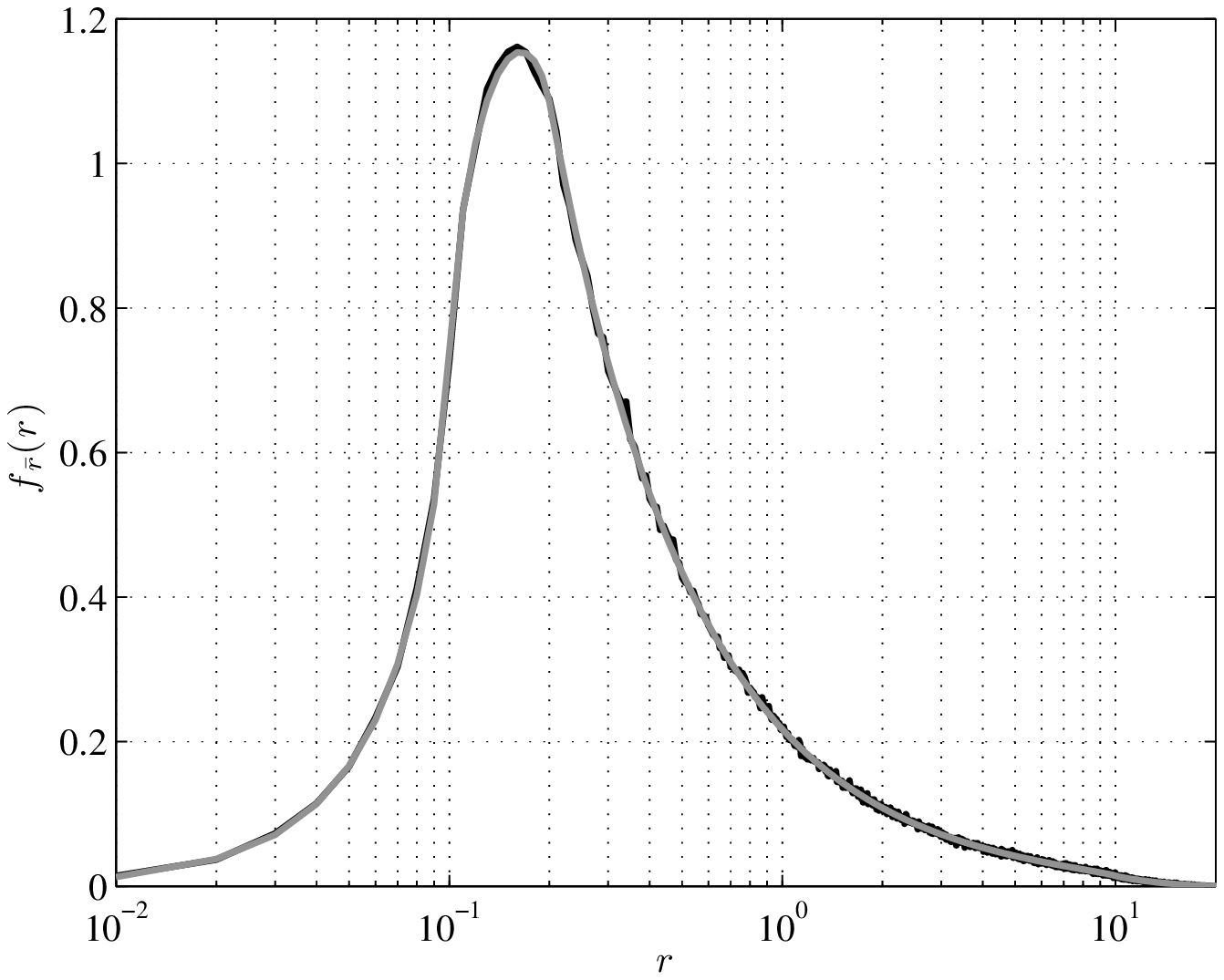}\label{fig:compDistsc}}
 \caption[]{ Comparison of empirical (black line) and derived (gray line) distributions assuming a uniform distribution for $\bar e$ and uniform and log distributions for $\bar a$.  For uniformly distributed semi-major axis we assumed $a \in [0.5, 1.5]$, leading to $r \in (0, 3]$, and for $\bar a$ uniform in $\log a$, we assumed $a \in [0.1, 10]$ leading to $r \in (0, 20]$. \label{fig:compDists}}
\end{figure}

To check these equations, we generate one million IID samples each from the uniform distributions in equations (\ref{eq:euniformpdf}), (\ref{eq:auniformpdf}) and (\ref{eq:Mdist}). \citep{press1992numerical}  From these, we calculate the eccentric anomalies via Newton-Raphson iteration applied to equation (\ref{eq:EtoM}), and then calculate $\nu$ and $r$ via equations (\ref{eq:nutoE}) and (\ref{eq:rdef}), respectively.  The sample PDFs are calculated  for these two parameters and compared with the results of our closed-form solutions, as shown in Figures \ref{fig:compDistsa} and \ref{fig:compDistsb}.  These plots demonstrate excellent agreement, validating the equations. 

While uniform distributions are a useful check, the semi-major axes of currently found planets appear more likely to be logarithmically distributed. \citep{currie2009semimajor} As an additional test, we will assume that $\bar a$ is uniformly distributed in $\log_{10}(a)$ for $a \in [10^{c_{min}}, 10^{c_{max}}]$ so that its PDF is:
\begin{equation}\label{eq:alogpdf}
f_{\bar a}(a) = \left\{
    \begin{array}{l l}
    \left(\Delta c\log(10) a\right)^{-1} &a \; \in \;  [10^{c_{min}}, 10^{c_{max}}]\\
    0 & \mathrm{else}
    \end{array}
        \right.
\end{equation}
where $\Delta c = c_{max} - c_{min}$.
If we retain the same uniform distribution for eccentricity, the true anomaly distribution remains the same, but the orbital radius density becomes:
\begin{align}
f_{\bar{r}}(r) =\frac{1}{2\pi \log(10)\Delta c a^2 r} & \left[
aC_1\sqrt{r} - a^2\log\left(a - r\right) - 2ia^2\log\left(\frac{32 r^3}{a}\left(r - a iC_1\sqrt{r}\right)\right)\right. \nonumber\\
&\left.\left. + 2a^2\log\left(-4\left(C_1r^{\frac{3}{2}} + r^2\right)\right) + 2r^2\log\left(\frac{C_2 - \sqrt{r-a}}{C_1 + \sqrt{r}}\right)\right]\right|_{10^{c_{min}}}^{10^{c_{max}}}\label{eq:rpdflog}
\end{align}
where $C_1$ and $C_2$ are defined as in equation (\ref{eq:rpdfuniform}).  We repeat the simulation, this time generating our sample of semi-major axes by exponentiating 10 with a sample of one million uniformly distributed random values between -1 and 1 (thereby setting $\Delta c$ to 2), and re-calculating the PDF for $f_{\bar r}$.  This is compared with the results from equation (\ref{eq:rpdflog}) in Figure \ref{fig:compDistsc}, and shows excellent agreement.

\begin{figure}[ht]
\centering
\subfigure[Uniform $f_{\bar{e}}$]{\includegraphics[width=0.45\columnwidth]{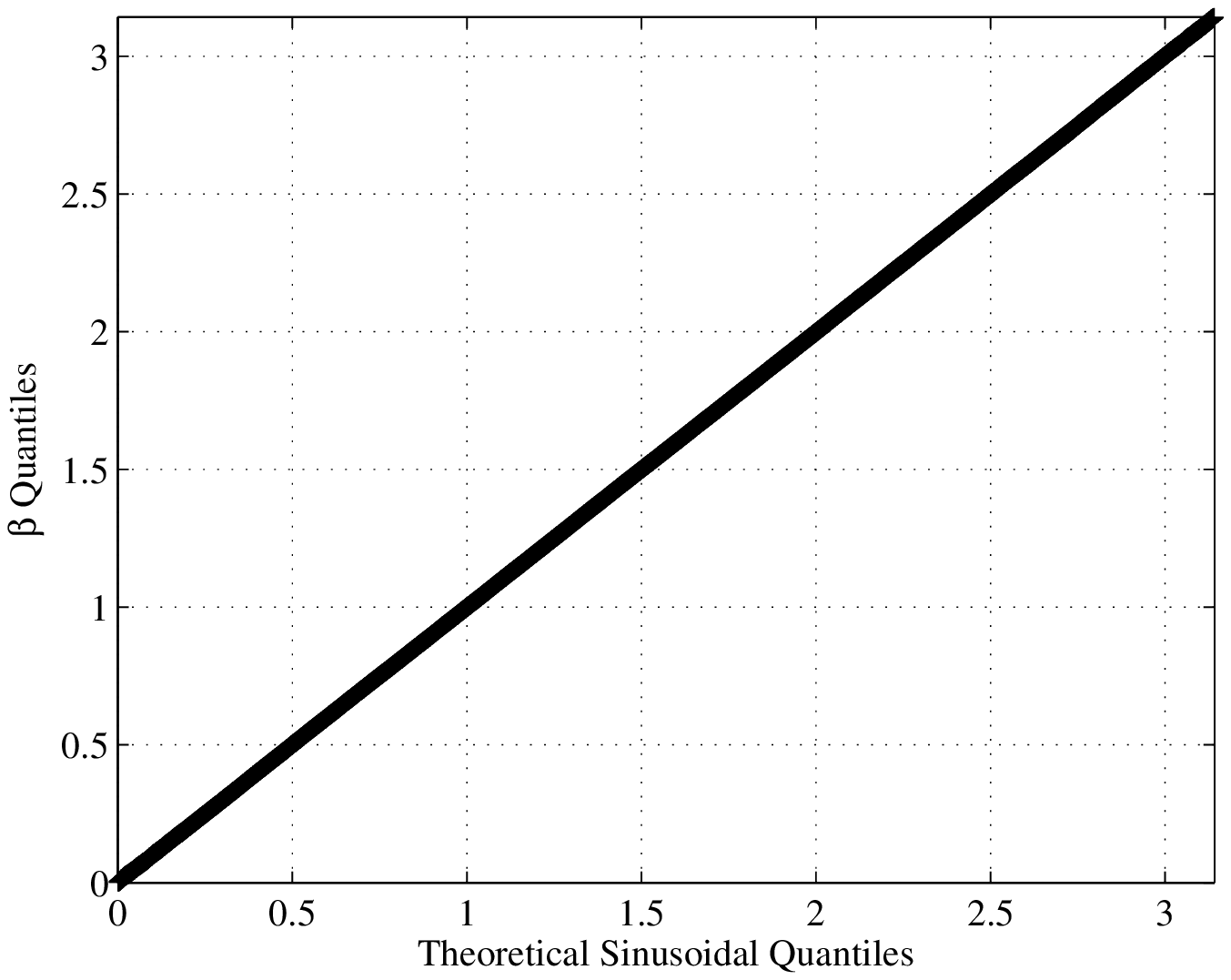}}
\subfigure[Step $f_{\bar{e}}$]{\includegraphics[width=0.45\columnwidth]{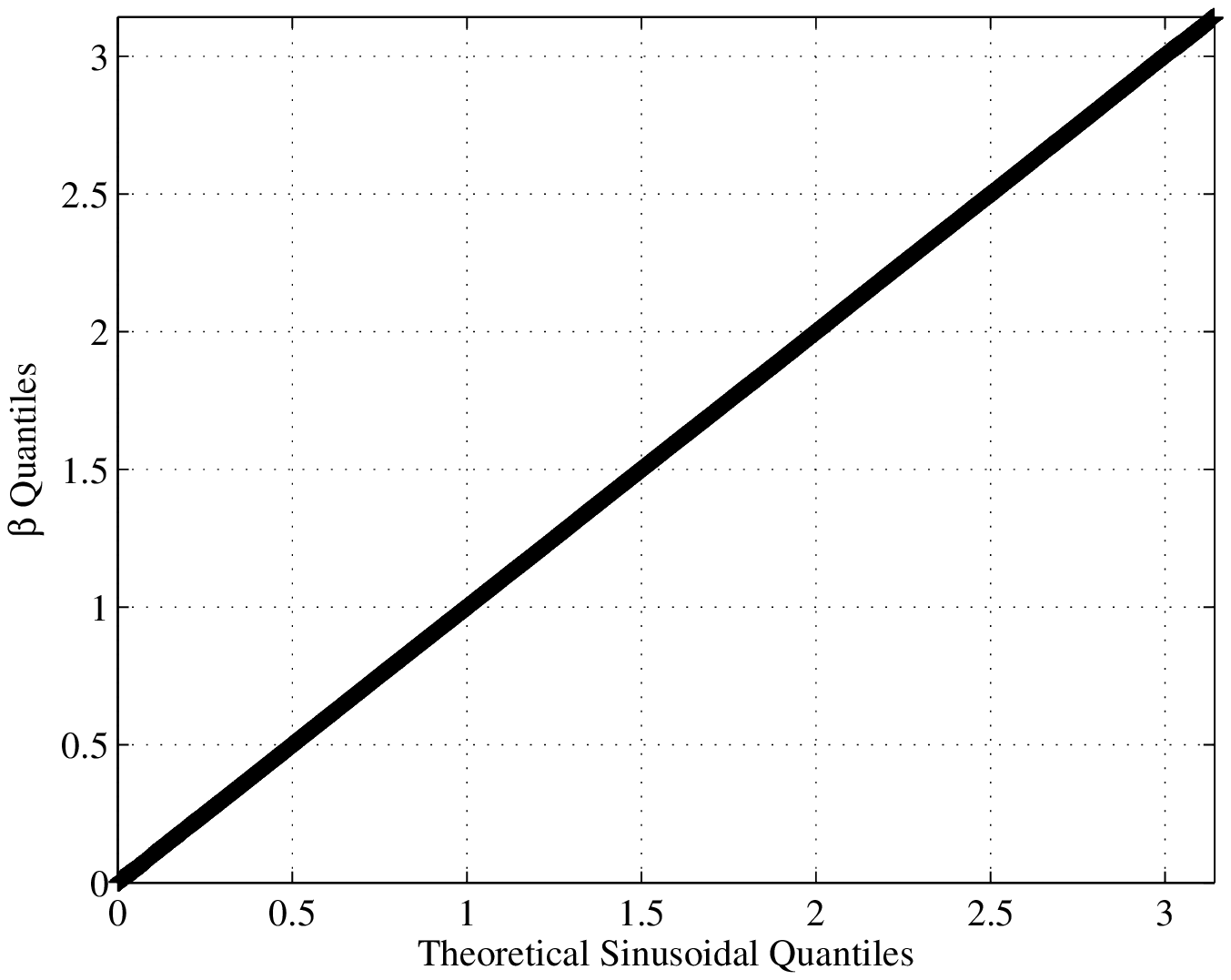}}
 \caption[]{ Q-Q plots of empirically derived $f_{\bar\beta}$ distributions compared with theoretical sinusoidal quantiles. \label{fig:qqplots}}
\end{figure}

We can also empirically check the assertion that the distribution of $\beta$ is independent of the distribution of $\bar e$.  We now generate two sets of IID samples, one using the uniform distribution of $\bar e$, and one distributed via a step function,
\begin{equation}
f_{\bar e}(e) = \left\{
    \begin{array}{l l}
   1.3 & e \; \in \; [0, 0.7) \\
   0.3 & e \; \in \; [0.7, 1]\\
    0 & \mathrm{else}
    \end{array}
    \right.
\end{equation}
where accurate sampling is achieved via simple rejection sampling.  The two sample sets are used to generate two different sample distributions of $\bar\beta$.  Figure \ref{fig:qqplots} shows Q-Q plots comparing the quantiles of these two samples to the theoretical quantiles of a sinusoidal distribution.  Quantiles for the ideal sinusoidal distribution are calculated by evaluating its inverted CDF at regularly spaced intervals, while quantiles of the two simulated distributions are calculated by ordering and binning the generated values.  As the resulting Q-Q plots follow the 45$^\circ$ line, this demonstrates that the simulated distributions are identical to the theoretical sinusoidal distribution. \citep{gibbons2003nonparametric}

\section{Conclusions}
Starting with a uniform distribution of orbital orientations and arbitrary PDFs for eccentricity and semi-major axis, we have derived expressions for the distributions of the remaining Keplerian orbital elements, and of parameters used to describe direct exoplanet detections.  We have demonstrated the independence of phase angle from the distributions of the Keplerian orbital elements, allowing for the calculation of the distributions of both apparent separation and flux ratio, which can be directly sampled to generate the completeness distribution.  At the same time, we have provided fully analytic forms for the distributions of Keplerian orbital elements based on uniform distributions of eccentricity and semi-major axis, as well as logarithmically distributed semi-major axes, assumptions often made in research related to planet-finding mission planning and analysis.  These forms should allow researchers to increase the efficiency of their simulations, and to empirically check for global errors generated by under-sampling.  The same equations and procedures described here will also be useful for other statistical work associated with planet-finding, including calculating the likelihood of transits or inferring the true distribution of orbital parameters from doppler spectroscopy surveys.
	
\acknowledgements{The authors gratefully acknowledge the input of Robert Vanderbei, whose comments and suggestions were invaluable, and Robert Brown, whose work has so much informed our own.}

\bibliographystyle{plainnat}
\bibliography{Main} 

\end{document}